\documentclass[11pt,a4paper,twosides]{article}

\input{Qcircuit}
\usepackage{amssymb}
\usepackage{url}
\usepackage{comment}

\usepackage[margin=4cm]{geometry}

\newcommand{\braket}[2]{\left\langle{#1}\vert{#2}\right\rangle}

\newcommand{\baregate}[1]{*{\xy *+<.6em>{#1};p\save+LU;+RU **\dir{-}\restore\save+RU;+RD **\dir{-}\restore\save+RD;+LD **\dir{-}\restore\POS+LD;+LU **\dir{-}\endxy}}

\newcommand{\multigateh}[2]{*+<0em,.6em>{\phantom{N}} \qw \POS[0,0].[0,#1];p !C *{#2},p \save+LU;+RU **\dir{-}\restore\save+RU;+RD **\dir{-}\restore\save+RD;+LD **\dir{-}\restore\save+LD;+LU **\dir{-}\restore}

\newcommand{\ghosth}{*+<0em,.6em>{\phantom{N}}}

\title{The sum-over-histories formulation of quantum computing}
\author{Ben Rudiak-Gould\footnote{\texttt{br276@cl.cam.ac.uk}}}

\begin{document}

\maketitle

\begin{abstract}

Since Deutsch (1985), quantum computers have been modeled exclusively
in the language of state vectors and the Schr\"odinger equation. We
present a complementary view of quantum circuits inspired by the path
integral formalism of quantum mechanics, and examine its application
to some simple textbook problems.

\end{abstract}

\section{Introduction}

At the undergraduate level, quantum mechanics is usually taught in the
so-called \textit{canonical} or \textit{Hamiltonian} formalism. In
this formalism the state of a quantum system is given by a vector in a
complex Hilbert space which evolves over time according to the
Schr\"odinger equation, and measurable properties are modeled by
Hermitian operators. The \textit{path integral} or \textit{Lagrangian}
formalism is generally introduced at the graduate level in a first
course in quantum field theory, though it is just as applicable to
nonrelativistic quantum mechanics. In the path integral formalism
there is no representation of the state of a quantum system between
preparation and measurement. The probability of a transition between
classical states is given by the integral, over every conceivable
intermediate classical history of the system, of a complex-valued
function of the history.

The canonical and path-integral approaches offer complementary views
of the same quantum theory. Often problems which are very difficult to
solve in one formalism are easy in the other: the path integral
formalism is far better suited to calculating scattering amplitudes in
high energy physics (with the help of Feynman diagrams), while the
canonical formalism is far better for calculating energy levels of
bound systems like the hydrogen atom.

Previous research in quantum computing seems to have used the
canonical formalism exclusively. In this paper we attempt to remedy
that oversight, by presenting a complementary view of quantum circuits
inspired by the path integral formalism, and examining its application
to some simple textbook problems.

\section{From quantum mechanics to quantum circuits}

\subsection{Discretizing the canonical formalism}

First we briefly review how the canonical formalism of continuum
quantum mechanics is related to the discrete quantum circuit model. We
write $\ket{\psi}$ for the state of a quantum system and say that it
evolves in time according to the Schr\"odinger equation,
$-i\hbar{\partial\over\partial t}\ket{\psi} = H(t)\ket{\psi}$.
Anticipating quantum circuits, we take the Hamiltonian $H$ to be a
function of time in order to model real-time classical control of the
quantum system.

We first discretize the phase space by decreeing that our quantum
system may only be in finitely many classical states. In the case of a
quantum computer with $n$ qubits there are $2^n$ states\footnote{This
is also true of classical computers, of course.}; then $\ket{\psi}$
may be seen as a vector in a $2^n$-dimensional space, or, with respect
to the computational basis, as a $2^n\times1$ column vector, each of
whose components is a complex amplitude. The operator $H$ becomes,
with respect to the same basis, a $2^n\times2^n$ matrix.

Next we discretize time, by supposing that the time between
preparation and measurement is divided into finitely many intervals of
length $\delta t$, and within each of these intervals $H$ does not
change. Then we can integrate the Schr\"odinger equation between $t$
and $t+\delta t$, turning it into a difference equation,
\begin{displaymath}
  \ket{\psi_{t+\delta t}} = e^{{i\over\hbar} H(t) \delta t} \ket{\psi_t}.
\end{displaymath}
It is easy to show that if $H$ is Hermitian then $e^{iH}$ is unitary;
so $U = e^{-{i\over\hbar} H(t) \delta t}$ is the unitary transition
matrix that appears in the usual model of quantum circuits.

\subsection{Discretizing the path-integral formalism}
\label{soh}

In the path-integral formalism, the quantum amplitude of a transition
from a classical initial state at time $t_i$ to a classical final
state at time $t_f$ is given by
\begin{displaymath}
  A = \int \mathrm{D}\phi \; e^{{i\over\hbar}S(\phi)},
  \qquad \mbox{where} \qquad
  S(\phi) = \int_{t_i}^{t_f} \mathrm{d}t \; L(\phi(t),\dot{\phi}(t),t).
\end{displaymath}
Here $\phi$ denotes a \textit{history}, i.e. any conceivable sequence
of classical states the system might occupy between $t_i$ and $t_f$.
The corresponding integral ranges over every possible history of the
system, whether or not that history is permitted by classical laws of
physics. The quantity $S(\phi)$ is the classical \textit{action}
associated with a particular history of the system, and $L$ is the
classical \textit{Lagrangian}, which is a function of the state
$\phi(t)$ and the rate of change of the state $\dot{\phi}(t)$ at a
given time. Like the Hamiltonian, and for the same reason, we also
give the Lagrangian an explicit time dependence. The relationship
between the action and the Lagrangian survives unchanged from
classical physics; the new physics is contained in the outer integral,
which states that each history, even if classically absurd,
contributes equally to the overall transition amplitude.

As before we discretize the path-integral formalism by limiting our
system to $2^n$ classical states and dividing time into intervals of
length $\delta t$. The histories $\phi$ become ordered tuples of
states; each state requires $n$ bits to describe, and each tuple
contains $\mathrm{\Delta}t/\delta t$ states (where $\mathrm{\Delta}t =
t_f - t_i$), so the outer integral becomes a sum over a discrete space
of $2^{n(\mathrm{\Delta}t/\delta t)}$ histories. The discretized
Lagrangian depends on the current state and the difference between the
current state and the next state, or equivalently on the current and
next states directly. So we have (with suitable normalization) $A =
\sum_{\phi} \exp\left({{i\over\hbar} \sum_{t=t_i}^{t_f}
L(\phi_t,\phi_{t+\delta t},t)}\right)$. Finally we move the
exponential inside the sum, obtaining
\begin{displaymath}
  A = \sum_{\phi} \prod_{t=t_i}^{t_f} B(\phi_t,\phi_{t+\delta t},t)
\end{displaymath}
where $B(\cdots) = e^{{i\over\hbar}L(\cdots)}$.

We will refer to the discretized path-integral formalism as the
\textit{sum-over-histories} formalism.

\subsection{Equivalence of the canonical and sum-over-histories formalisms}
\label{equiv}

There is a delightfully simple way to see that the discrete canonical
and sum-over-histories formalisms are mathematically equivalent. In
the canonical formalism we start with a system prepared in the state
$\ket{\psi_0}$, which we may assume to be an eigenstate of the
computational basis. To this we apply a sequence of unitary
transformations, obtaining $\ket{\psi_1} = U^{(1)}\ket{\psi_0},
\ldots, \ket{\psi_m} = U^{(m)}\ket{\psi_{m-1}}$. Finally we measure
all of the qubits, obtaining the computational eigenstate
$\ket{\psi_f}$ with probability
$\left\vert\braket{\psi_f}{\psi_m}\right\vert^2$. Combining these
steps into one, we have that the probability is the squared modulus of
$A = \bra{\psi_f} U^{(m)}\cdots U^{(1)} \ket{\psi_0}.$ Writing out the
matrix product explicitly,
\begin{displaymath}
  A \quad = \quad
    \sum_{i_1}\cdots\sum_{i_{m-1}} U^{(m)}_{i_m i_{m-1}} \cdots
    U^{(2)}_{i_2 i_1} U^{(1)}_{i_1 i_0}
  \quad = \quad
    \sum_{i_1}\cdots\sum_{i_{m-1}} \prod_{j=1}^m U^{(j)}_{i_j i_{j-1}}
\end{displaymath}
where each sum is taken over the $2^n$ states of the computational
basis, and $i_0$ and $i_m$ are the initial and final basis states
$\ket{\psi_0}$ and $\ket{\psi_f}$. But this is exactly the sum over
histories, with $\phi = (i_1,\ldots,i_{m-1}) \in (2^n)^{m-1}$ and
$B(\phi_{j-1}, \phi_j, t) = U^{(j)}_{i_{j}i_{j-1}}$.\footnote{In fact
we have glossed over a serious problem here: the derivation of section
\ref{soh} implies that the modulus of $B$ is independent of its
arguments, and this certainly is not true of the components of the
matrices $U$. To actually obtain arbitrary unitary matrices from the
path-integral formalism we must further subdivide $\delta t$. But for
the purposes of this section it is enough to simply generalize $B$.}

\section{Quantum circuits in the sum-over-histories picture}

The reader may be wondering at this point why we bothered to introduce
the sum-over-histories formalism, if its difference from the canonical
formalism amounts to mere algebra. In this section we justify its
existence by showing that it leads to an interesting new view of
quantum circuits.

Returning to the canonical formalism for the moment, consider a small
quantum system of three qubits (call them $a$, $b$ and $c$) to which
we successively apply three unitary transformations, transforming the
system through two intermediate states to a final state. We might
illustrate this as follows:

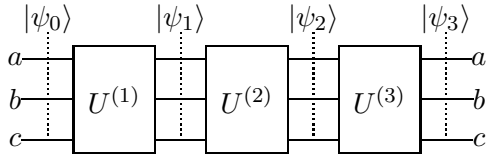
\begin{displaymath}
  \begin{array}{c}
  \Qcircuit @C=0em @R=0.4em @!R {
    & \push{\ket{\psi_0}} \ar @{.} [ddd] &
    & \push{\ket{\psi_1}} \ar @{.} [ddd] &
    & \push{\ket{\psi_2}} \ar @{.} [ddd] &
    & \push{\ket{\psi_3}} \ar @{.} [ddd] \\
    \push{a} & \qw & \multigate{2}{U^{(1)}} & \qw & \multigate{2}{U^{(2)}} & \qw & \multigate{2}{U^{(3)}} & \qw & \push{a} \qw \\
    \push{b} & \qw & \ghost{U^{(1)}}        & \qw & \ghost{U^{(2)}}        & \qw & \ghost{U^{(3)}}        & \qw & \push{b} \qw \\
    \push{c} & \qw & \ghost{U^{(1)}}        & \qw & \ghost{U^{(2)}}        & \qw & \ghost{U^{(3)}}        & \qw & \push{c} \qw
  }
  \end{array}
\end{displaymath}

In the sum-over-histories formalism we never deal with general state
vectors, only with classical states---i.e. computational basis
vectors---so instead of $\ket{\psi_0}, \ldots, \ket{\psi_3}$ we may as
well write $\ket{a_0b_0c_0},$ $\ldots,$ $\ket{a_3b_3c_3}$, with each
$a_k, b_k, c_k$ taking on a value from $\{0,1\}$ in a particular
history. In fact, there is no need to write these variables in a ket
above the circuit: it is clearer to place them on the wires
themselves.
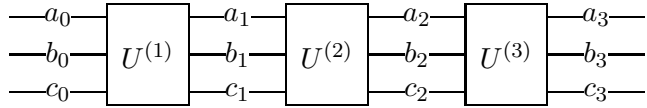
\begin{displaymath}
  \begin{array}{c}
  \Qcircuit @C=1.2em @R=0.4em @!R {
    & \push{a_0}\qw & \multigate{2}{U^{(1)}} & \push{a_1}\qw & \multigate{2}{U^{(2)}} & \push{a_2}\qw & \multigate{2}{U^{(3)}} & \push{a_3}\qw & \qw \\
    & \push{b_0}\qw & \ghost{U^{(1)}}        & \push{b_1}\qw & \ghost{U^{(2)}}        & \push{b_2}\qw & \ghost{U^{(3)}}        & \push{b_3}\qw & \qw \\
    & \push{c_0}\qw & \ghost{U^{(1)}}        & \push{c_1}\qw & \ghost{U^{(2)}}        & \push{c_2}\qw & \ghost{U^{(3)}}        & \push{c_3}\qw & \qw \\
  }
  \end{array}
\end{displaymath}

So if the system is prepared in the state $i_0 = \ket{a_0b_0c_0}$, the
amplitude that it will be found in the state $i_3 = \ket{a_3b_3c_3}$
after application of these three unitary gates can be found by summing
the contribution of the $2^6=64$ histories arising from all possible
choices of $a_1,b_1,c_1,a_2,b_2,c_2 \in \{0,1\}$. The contribution of
each history is the product $U^{(1)}_{\ket{a_1b_1c_1}\ket{a_0b_0c_0}}
U^{(2)}_{\ket{a_2b_2c_2}\ket{a_1b_1c_1}}
U^{(3)}_{\ket{a_3b_3c_3}\ket{a_2b_2c_2}}$. Clearly this extends to any
number of gates; it merely paraphrases the explicit formula of Section
\ref{soh}.

Suppose now that $U^{(2)}$ acts only on qubit $b$---i.e. that $U^{(2)}
= I \otimes Q \otimes I$, where $I = \left({1\,0 \atop 0\,1}\right)$
and $Q$ is some $2\times2$ unitary matrix.
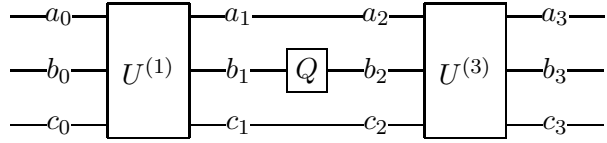
\begin{displaymath}
  \begin{array}{c}
  \Qcircuit @C=1.2em @R=0.4em @!R {
    & \push{a_0}\qw & \multigate{2}{U^{(1)}} & \push{a_1}\qw & \qw      & \push{a_2}\qw & \multigate{2}{U^{(3)}} & \push{a_3}\qw & \qw \\
    & \push{b_0}\qw & \ghost{U^{(1)}}        & \push{b_1}\qw & \gate{Q} & \push{b_2}\qw & \ghost{U^{(3)}}        & \push{b_3}\qw & \qw \\
    & \push{c_0}\qw & \ghost{U^{(1)}}        & \push{c_1}\qw & \qw      & \push{c_2}\qw & \ghost{U^{(3)}}        & \push{c_3}\qw & \qw \\
  }
  \end{array}
\end{displaymath}

Most of the entries of $U^{(2)}$ are now zero; specifically,
$U^{(2)}_{\ket{a_2b_2c_2}\ket{a_1b_1c_1}} = I_{a_2a_1} Q_{b_2b_1}
I_{c_2c_1}$, which is zero at least when $a_1 \ne a_2$ or $c_1 \ne
c_2$. Therefore, the total contribution of any history with $a_1 \ne
a_2$ or $c_1 \ne c_2$ will be zero. Therefore, we need not consider
those histories at all! We can easily exclude them by taking $a_1 =
a_2 = a_{12}$ and $c_1 = c_2 = c_{12}$.
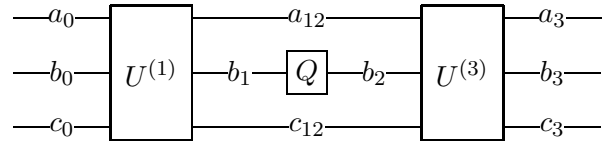
\begin{displaymath}
  \begin{array}{c}
  \Qcircuit @C=1.2em @R=0.4em @!R {
    & \push{a_0}\qw & \multigate{2}{U^{(1)}} & \qw           & \push{a_{12}}\qw & \qw           & \multigate{2}{U^{(3)}} & \push{a_3}\qw & \qw \\
    & \push{b_0}\qw & \ghost{U^{(1)}}        & \push{b_1}\qw & \gate{Q}         & \push{b_2}\qw & \ghost{U^{(3)}}        & \push{b_3}\qw & \qw \\
    & \push{c_0}\qw & \ghost{U^{(1)}}        & \qw           & \push{c_{12}}\qw & \qw           & \ghost{U^{(3)}}        & \push{c_3}\qw & \qw \\
  }
  \end{array}
\end{displaymath}

Now we may compute the same transition amplitude as before by summing
the contribution of just $16$ histories arising from all possible
choices of $a_{12},b_1,b_2,c_{12} \in \{0,1\}$.

Generalizing this idea, we arrive at the following formulation of
sum-over-histories for quantum circuits. We say that an
\textit{internal wire} begins as the output of a gate and ends as the
input to another gate, possibly controlling other gates along the way.
For example, this is an internal wire:

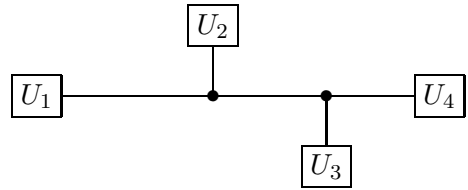
\begin{displaymath}
  \begin{array}{c}
  \Qcircuit @R=1em {
    & & \baregate{U_2} & & \\
    \baregate{U_1} & \qw & \ctrl{-1} & \ctrl{1} & \gate{U_4} \\
    & & & \baregate{U_3} & \\
  }
  \end{array}
\end{displaymath}

This is three internal wires (one above, two below):

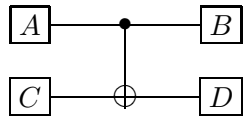
\begin{displaymath}
  \begin{array}{c}
  \Qcircuit @R=1.2em {
    \baregate{A} & \ctrl{1} & \gate{B} \\
    \baregate{C} & \targ & \gate{D}
  }
  \end{array}
\end{displaymath}

An \textit{external wire} begins as an input to the circuit and ends
at a gate, or begins at a gate and ends as an output from the circuit,
or begins as an input and ends as an output. Like an internal wire, it
may control other gates along the way.

We assign a value of either $0$ or $1$ to each internal and external
wire in every possible way. The values of the external wires are fixed
by the specification of the inputs and outputs of the circuit, while
the values of the internal wires are not; thus there are $2^w$
possible assignments (histories), where $w$ is the number of internal
wires. For each of these histories, for each gate in the circuit, we
find the element of the unitary matrix indexed by the inputs and
outputs of the gate. We multiply these to find the contribution of the
history, and add the contributions of every history to find the
transition amplitude, whose squared modulus is the transition
probability.

Note that even though our new history-enumeration rule eliminates all
zeroes that arise from tensor products with the identity, there may
still be zeroes lurking within our gates. We will say that a gate
\textit{rejects} a history if it causes that history's contribution to
go to zero. A gate defined by a matrix containing only ones and zeroes
will either reject a particular history or contribute a factor of one,
effectively doing nothing (which we will call \textit{accepting} the
history). A gate which either accepts or rejects every history is
\textit{classical}.

\subsection{Topological quantum computing}
\label{topo}

Returning to the canonical formalism, consider the circuit
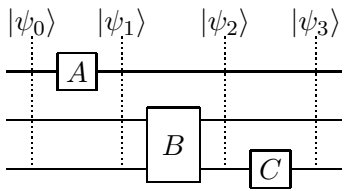
\begin{displaymath}
  \begin{array}{c}
  \Qcircuit @C=0em @R=0.4em @!R {
    & \push{\ket{\psi_0}} \ar @{.} [ddd] &
    & \push{\ket{\psi_1}} \ar @{.} [ddd] &
    & \push{\ket{\psi_2}} \ar @{.} [ddd] &
    & \push{\ket{\psi_3}} \ar @{.} [ddd] \\
    & \qw & \gate{A} & \qw & \qw & \qw & \qw & \qw & \qw \\
    & \qw & \qw & \qw & \multigate{1}{B} & \qw & \qw & \qw & \qw \\
    & \qw & \qw & \qw & \ghost{B} & \qw & \gate{C} & \qw & \qw
  }
  \end{array}
\end{displaymath}
It is of course true (though not immediately obvious in this
formalism) that we may switch the order of gates $A$ and $B$ in this
circuit without affecting the subsequent states ($\ket{\psi_2}$ and
$\ket{\psi_3}$). But state $\ket{\psi_1}$ does change: in the revised
circuit it is replaced by an entirely different state (call it
$\ket{\psi_1'}$). There is no straightforward relationship between
$\ket{\psi_1}$ and $\ket{\psi_1'}$.

If we apply sum-over-histories to this circuit, it is immediately
clear that $A$ and $B$ may be switched without affecting the result;
it is equally clear that \emph{nothing changes} when we switch them,
not even unobservable bookkeeping state.

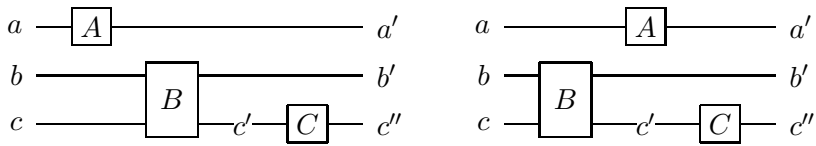
\begin{displaymath}
  \begin{array}{c}
  \Qcircuit @C=1.2em @R=0.4em @!R {
    \lstick{a} & \gate{A} & \qw              &          \qw & \qw      & \rstick{a'}\qw \\
    \lstick{b} & \qw      & \multigate{1}{B} &          \qw & \qw      & \rstick{b'}\qw \\
    \lstick{c} & \qw      & \ghost{B}        & \push{c'}\qw & \gate{C} & \rstick{c''}\qw
  }
  \end{array}
  \qquad\qquad
  \begin{array}{c}
  \Qcircuit @C=1.2em @R=0.4em @!R {
    \lstick{a} & \qw              & \gate{A}     & \qw      & \rstick{a'}\qw \\
    \lstick{b} & \multigate{1}{B} &          \qw & \qw      & \rstick{b'}\qw \\
    \lstick{c} & \ghost{B}        & \push{c'}\qw & \gate{C} & \rstick{c''}\qw
  }
  \end{array}
\end{displaymath}

Circuits are in some sense topological in the sum-over-histories
picture. We do not have any state which cuts across all wires,
imposing a linear order on our gates. Indeed, it is not clear that we
cannot use sum-over-histories to find a transition amplitude for a
circuit like

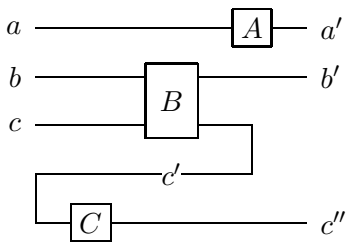
\begin{displaymath}
  \begin{array}{c}\Qcircuit @C=1.2em @R=0.4em @!R {
    \lstick{a} &      \qw & \qw              & \gate{A}   & \rstick{a'}\qw \\
    \lstick{b} &      \qw & \multigate{1}{B} & \qw        & \rstick{b'}\qw \\
    \lstick{c} &      \qw & \ghost{B}        & \qw\qwx[1] & \\
               &      \qw & \push{c'}\qw     &        \qw & \\
      \qwx[-1] & \gate{C} &              \qw &        \qw & \rstick{c''}\qw
  }\end{array}
\end{displaymath}
which has no obvious interpretation at all in the canonical formalism.

Some care is necessary here because many gates are not symmetric with
respect to rearrangement of their parameters. If we were to move the
$C$ gate to the location of the $c'$ wire label, we would have to
indicate somehow that its left and right parameters had switched
places. An obvious solution is to write $C^t$ instead of $C$ in this
situation (note that this is the transpose, not the adjoint).
Fortunately, many common gates turn out to be perfectly symmetric, and
here we will look at three interesting examples.

First, the CNOT gate, which we now view as a three-parameter gate,
accepts every history where the sum of its parameters is even, and
rejects all other histories. So it is in fact perfectly symmetric
between ``input,'' ``output,'' and ``control,'' and we need not
distinguish them at all. We will call the symmetric version of CNOT
the \textit{xor gate}.

Second, diagonal gates like $Z$ and $\pi/8$, which we will here call
\textit{phase gates}, are perfectly symmetric between input and
output. In fact, the input and output are actually the same wire: the
off-diagonal zeroes cause rejection of any history in which they carry
different values. Another way to see this is to look at the sequence
controlled-controlled-Z, controlled-Z, Z:
\begin{displaymath}
  \left(\begin{array}{cccccccc}
    1&0&0&0&0&0&0&0\\
    0&1&0&0&0&0&0&0\\
    0&0&1&0&0&0&0&0\\
    0&0&0&1&0&0&0&0\\
    0&0&0&0&1&0&0&0\\
    0&0&0&0&0&1&0&0\\
    0&0&0&0&0&0&1&0\\
    0&0&0&0&0&0&0&-1
  \end{array}\right)
  ,
  \left(\begin{array}{cccc}
    1&0&0&0\\
    0&1&0&0\\
    0&0&1&0\\
    0&0&0&-1
  \end{array}\right)
  ,
  \left(\begin{array}{cc}
    1&0\\
    0&-1
  \end{array}\right)
\end{displaymath}
It is clear that this sequence has been cut off one term short of its
natural end. The next term ought to be $(-1)$, a unitary $1\times1$
\emph{zero}-qubit gate which we will call the \textit{--1 gate}. The
same applies to any single-qubit gate described by a diagonal matrix.
These gates actually operate on no qubits at all; for consistency we
should write
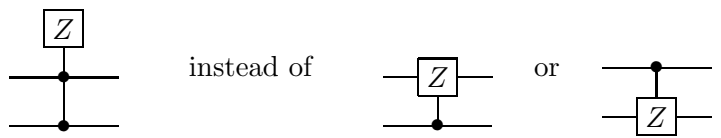
\begin{displaymath}
  \begin{array}{c}
  \Qcircuit @C=1.2em @R=0.4em @!R {
    & \baregate{Z} & \\
    & \ctrl{-1}    & \qw \\
    & \ctrl{-2}    & \qw
  }
  \end{array}
  \qquad\mbox{instead of}\qquad
  \begin{array}{c}
  \Qcircuit @C=1.2em @R=0.4em @!R {
    &           & \\
    & \gate{Z}  & \qw \\
    & \ctrl{-1} & \qw
  }
  \end{array}
  \quad\mbox{or}\quad
  \begin{array}{c}
  \Qcircuit @C=1.2em @R=0.4em @!R {
    &           & \\
    & \ctrl{1} & \qw \\
    & \gate{Z} & \qw
  }
  \end{array}.
\end{displaymath}
A free-floating (uncontrolled) phase gate contributes the same phase
to every history, which has no observable effect.

Third, the uncontrolled Hadamard gate does not reject any combination
of input and output; if both of its parameters are $1$ it contributes
a factor of $-1$ to the history; and regardless of its parameters it
contributes an additional factor of $1\over\sqrt2$. We can absorb the
factor of $1\over\sqrt2$ into a post-normalization step, since it is
the same for every history. Without this factor, the effect of the
Hadamard gate is exactly that of the controlled-Z gate. So the
Hadamard gate is nothing but a doubly-controlled $-1$ gate which just
happens to be written with its control wires on the left and right
instead of the top and bottom.

It is well known that classical gates plus $H$, $Z$ and $\pi/8$
constitute a universal quantum gate set. We have just shown that in
the sum-over-histories formalism $H$, $Z$ and $\pi/8$ are all special
cases of the controlled zero-qubit phase gate. The extra power of
quantum computers seems to be vested entirely in gates that do not
affect any qubits!

\section{Case study: quantum teleportation}
\label{apps}

Here is a typical quantum teleportation circuit.
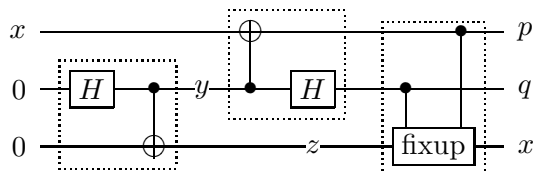
\begin{displaymath}
  \begin{array}{c}
  \Qcircuit @C=1em @R=0.7em @!R {
    \lstick{x} & \qw & \qw & \qw & \targ & \qw & \qw & \qw & \ctrl{2} & \rstick{p} \qw \\
    \lstick{0}    & \gate{H} & \ctrl{1} & \push{y} \qw & \ctrl{-1} & \gate{H} & \qw & \ctrl{1} & \qw & \rstick{q} \qw \\
    \lstick{0}    & \qw & \targ   & \qw & \qw & \push{z} \qw & \qw & \multigateh{1}{\mbox{fixup}} & \ghosth & \rstick{x} \qw
    \gategroup{2}{2}{3}{3}{.7em}{..}
    \gategroup{1}{5}{2}{6}{.7em}{..}
    \gategroup{1}{8}{3}{9}{.7em}{..}
  }
  \end{array}
\end{displaymath}
The boxes mark the three conceptually separate parts of this circuit;
left to right, they are the Bell state creation (performed jointly by
Alice and Bob), the Bell state uncreation (performed by Alice), and
the ``fixup'' step (performed by Bob with two bits of classical
information from Alice).

The question that interests us here is, what is the
$\begin{array}{c}\Qcircuit{\baregate{\mbox{fixup}}}\end{array}$?
Working this out from the state vector is tedious. In this section we
will see how to find the answer from the sum over histories without
any calculation at all.


We begin by replacing Hadamard gates with doubly-controlled $-1$ gates.

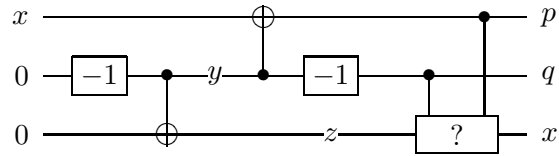
\begin{displaymath}
  \begin{array}{c}
  \Qcircuit @C=1em @R=0.7em @!R {
    \lstick{x} & \qw & \qw & \qw & \targ & \qw & \qw & \qw & \ctrl{2} & \rstick{p} \qw \\
    \lstick{0}    & \gate{-1} & \ctrl{1} & \push{y} \qw & \ctrl{-1} & \gate{-1} & \qw & \ctrl{1} & \qw & \rstick{q} \qw \\
    \lstick{0}    & \qw & \targ   & \qw & \qw & \push{z} \qw & \qw & \multigateh{1}{\mbox{?}} & \ghosth & \rstick{x} \qw
  }
  \end{array}
\end{displaymath}

One of the controls for the first $-1$ gate is always $0$, so the gate
will never be active. We can simply drop it:

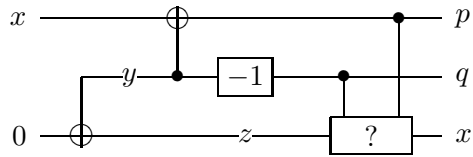
\begin{displaymath}
  \begin{array}{c}
  \Qcircuit @C=1em @R=0.7em @!R {
    \lstick{x} & \qw & \qw & \targ & \qw & \qw & \qw & \ctrl{2} & \rstick{p}\qw \\
                  & \qwx[1] & \push{y}\qw & \ctrl{-1} & \gate{-1} & \qw & \ctrl{1} & \qw & \rstick{q}\qw \\
    \lstick{0}    & \targ   & \qw & \qw & \push{z} \qw & \qw & \multigateh{1}{\mbox{?}} & \ghosth & \rstick{x} \qw
  }
  \end{array}
\end{displaymath}

At the lower left we have an xor gate one of whose parameters is
always zero. Under these circumstances, the gate will accept a circuit
iff its other two parameters are equal. So we may as well short those
parameters together: $y$ and $z$ are the same wire.

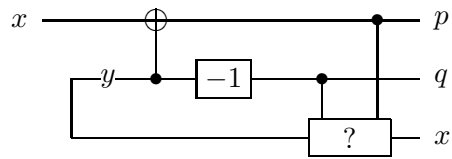
\begin{displaymath}
  \begin{array}{c}
  \Qcircuit @C=1em @R=0.7em @!R {
    \lstick{x} & \qw & \qw & \targ & \qw & \qw & \qw & \ctrl{2} & \rstick{p} \qw \\
                  & \qwx[1] & \push{y}\qw & \ctrl{-1} & \gate{-1} & \qw & \ctrl{1} & \qw & \rstick{q} \qw \\
                  & \qwx[-1] & \qw & \qw & \qw & \qw & \multigateh{1}{\mbox{?}} & \ghosth & \rstick{x} \qw
  }
  \end{array}
\end{displaymath}

A bit of rearrangement gives

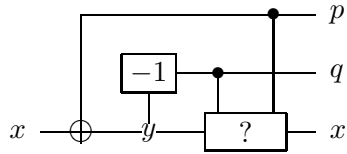
\begin{displaymath}
  \begin{array}{c}
  \Qcircuit @C=1em @R=0.7em @!R {
               & \qwx[2] & \qw             & \qw                      & \ctrl{2} & \rstick{p}\qw \\
               &         & \baregate{-1}   & \ctrl{1}                 & \qw      & \rstick{q}\qw \\
    \lstick{x} & \targ   & \push{y}\qw\qwx & \multigateh{1}{\mbox{?}} & \ghosth  & \rstick{x}\qw
  }
  \end{array}
\end{displaymath}

The situation, then, is as follows: if $p$, then $y=\neg x$, otherwise
$y=x$; and if $q\,\&\,y$, then the whole world has been rotated by
$180^\circ$ in the complex plane. The job of the $?$ is to undo the
rotation and recover $x$, and it is now easy to see how to do this.
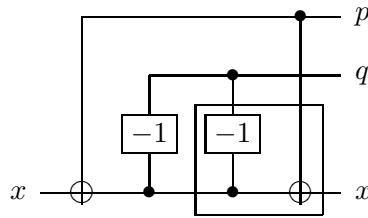
\begin{displaymath}
  \begin{array}{c}
  \Qcircuit @C=1em @R=0.7em @!R {
               & \qwx[3] & \qw & \qw & \ctrl{3} & \rstick{p} \qw \\
               & & \qwx[1] & \ctrl{1} & \qw & \rstick{q} \qw \\
               & & \baregate{-1} & \baregate{-1} & & \\
    \lstick{x} & \targ & \ctrl{-1} & \ctrl{-1} & \targ & \rstick{x} \qw
    \gategroup{3}{4}{4}{5}{.7em}{-}
  }
  \end{array}
\end{displaymath}
Backporting this result to the conventional notation gives us the
solution to the original problem.
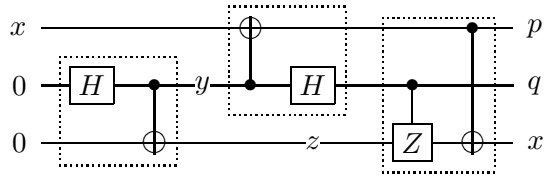
\begin{displaymath}
  \begin{array}{c}
  \Qcircuit @C=1em @R=0.7em @!R {
    \lstick{x} & \qw & \qw & \qw & \targ & \qw & \qw & \qw & \ctrl{2} & \rstick{p} \qw \\
    \lstick{0}    & \gate{H} & \ctrl{1} & \push{y} \qw & \ctrl{-1} & \gate{H} & \qw & \ctrl{1} & \qw & \rstick{q} \qw \\
    \lstick{0}    & \qw & \targ   & \qw & \qw & \push{z} \qw & \qw & \gate{Z} & \targ & \rstick{x} \qw
    \gategroup{2}{2}{3}{3}{.7em}{..}
    \gategroup{1}{5}{2}{6}{.7em}{..}
    \gategroup{1}{8}{3}{9}{.7em}{..}
  }
  \end{array}
\end{displaymath}

With a bit of practice it is possible to perform all of these steps
mentally, reading the answer directly off of the original circuit. As
an exercise, the reader might try to find the correct fixup step for
this closely related superdense coding circuit, and also label the
marked wires with $p$, $q$, $x$, $y$, and $z$ according to their
correspondence to the wires in the diagram above.

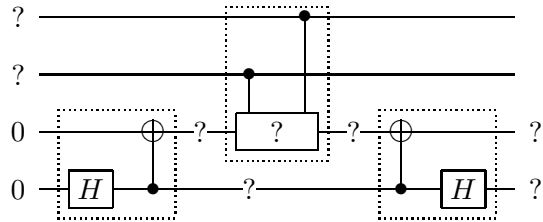
\begin{displaymath}
  \begin{array}{c}
  \Qcircuit @C=1em @R=0.7em @!R {
    \lstick{?} & \qw & \qw & \qw & \qw & \ctrl{2} & \qw & \qw & \qw & \qw \\
    \lstick{?} & \qw & \qw & \qw & \ctrl{1} & \qw & \qw & \qw & \qw & \qw \\
    \lstick{0} & \qw & \targ & \push{?}\qw & \multigateh{1}{\mbox{?}} & \ghosth & \push{?}\qw & \targ & \qw & \rstick{?} \qw \\
    \lstick{0} & \gate{H} & \ctrl{-1} & \qw & \push{?}\qw & \qw & \qw & \ctrl{-1} & \gate{H} & \rstick{?} \qw
    \gategroup{3}{2}{4}{3}{.7em}{..}
    \gategroup{1}{5}{3}{6}{.7em}{..}
    \gategroup{3}{8}{4}{9}{.7em}{..}
  }
  \end{array}
\end{displaymath}

\section{Conclusions}

Is the sum-over-histories formalism useful? Path-integral techniques
in physics have been fantastically successful, but it is not clear
that this is relevant to quantum computing. In physics one is often
interested in calculating quantum corrections to a roughly classical
quantity, while in quantum computing one generally wants to be as far
from classical behavior as possible. The QFT, the key to so many
quantum algorithms, is in a certain sense maximally nonclassical.
Probably there will be no polynomial-time circuit simulation
algorithms forthcoming from the sum-over-histories formalism.

Sum-over-histories certainly gives a useful perspective on some
results in complexity theory. The fact that $\mathbf{BQP} \subseteq
\mathbf{PSPACE}$, surprising in the canonical formalism, is almost
self-evident in sum-over-histories. A na\"{\i}ve circuit simulator
based on sum-over-histories requires very little space, especially
compared to the exorbitant requirements of a state-vector simulator.
Of course, all of this is well known already.

This author's particular interest is language design, and he has some
hope that sum-over-histories might be useful here. What properties
would we expect of a language which compiles to a ``topological''
quantum circuit, in the sense of Section \ref{topo}, instead of a
sequential circuit? A serious complication is that it is not clear how
to go backwards from a topological circuit to something actually
realizable on quantum hardware. It would be necessary to find an
algorithm which not only recovers a circuit, but recovers it quickly
enough that it is not faster to simply simulate the topological
circuit classically. Worse, only some topological circuits correspond
to any realizable circuit, and it is not clear how to characterize
these.

The author is particularly hopeful that sum-over-histories could be
useful in education. It is a truism that one has not understood
something until one has understood it in two different ways. Many
textbooks, as well as popular introductions to quantum mechanics,
foster the impression that the state vector is \emph{really} the state
of a quantum system, and the universe really is keeping track of $2^n$
complex amplitudes. In fact there are good theoretical reasons not to
take state vectors too seriously as a model of physical reality. It
might also be easier in the sum-over-histories picture to motivate the
connection between classical and quantum computing, and the embedding
of the former in the latter. The notion of classical gates
``accepting'' and ``rejecting'' histories is reasonably intuitive, and
a circuit containing only classical gates rejects all histories but
one.

\end{document}